\documentclass{cernphprepmod}   
\usepackage{relsize}
\usepackage{hyperref} 
\usepackage{lineno}
\begin{document}
\begin{titlepage}
\PHdate{8 February 2021} 
\vspace{1.2cm}
\title{\LARGE Space Charge Effects in Noble Liquid Calorimeters\\
 and Time Projection Chambers}
\begin{Authlist}
Sandro Palestini 
\Instfoot{a1}{CERN, 1211 Geneva 23, Switzerland}

\end{Authlist}
\ShortAuthor{}
\ShortTitle{}

\begin{abstract}
{The subject of space charge in ionization detectors is reviewed,  showing how the observations and the formalism used to describe the effects have evolved, starting with applications to calorimeters and reaching recent, large-size time projection chambers. 
General scaling laws, and different ways to present and model the effects are presented.
The relation between space-charge effects and the boundary conditions imposed on the side faces of the detector are discussed, 
together with a design solution that mitigates part of the effects.  
The implications of the relative size of drift length and transverse detector size are illustrated. %
Calibration methods are briefly discussed. } 
\end{abstract}
\Submitted{Submitted to Instruments\\
{\small Special Issue:~~Liquid Argon Detectors: Instrumentation and Applications}}

\end{titlepage}
\section{Introduction}
The term space charge is referred to the underlying distribution of charge, usually positive ions, reaching a level
capable to affect significantly the electric field established by the voltage on the electrodes of 
an electronic device.  
The change in electric field affects the motion of the main charge carriers, usually electrons, and therefore changes the signal being read out.
This subject was discussed first in the context of gas diodes~\cite{SpaceCharge-hystorical-1,SpaceCharge-hystorical-2}, before becoming 
relevant for particle detectors, namely in the context of 
drift chambers with long collection time~\cite{SpaceCharge-gasTPC-1,SpaceCharge-gasTPC-2}. 
In these cases, the positive ions are produced through the mechanism of gas multiplication in proximity of the anode.
Ionization detectors without gas multiplication can be affected by space charge as well, 
if the time necessary to dispose of the positive ions becomes long enough 
to cause significant accumulation of positive charge density.  
In particular, this situation occurs in detectors based on dense media, such as liquid argon or krypton, used in calorimetry for high radiation intensity applications, 
or for large-size time projection chambers (TPC), where even low levels of radiation, together with long drift paths, can cause significant space-charge effects. 

 This paper deals with ionization detectors without gas multiplication --- with the exception of a short discussion of liquid argon TPCs in which gas multiplication contributes to space charge.   Section~\ref{sec:history} contains first a historical review of relevant observations, 
sometimes reported together with developments of the formalism used to describe the effect. 
Space-charge effects occurring near the side faces of TPCs, and a design solution for their mitigation are discussed in Section~\ref{sec:models},
which also contains a comparison of the predictions of multidimensional models with observations.
Section~\ref{sec:dual-phase} contains the discussion of space charge in dual-phase detectors. 
Fluid motion represents a limit to the validity of electrostatic models for space-charge effects, and reinforces the need of calibration procedures. 
Section~\ref{sec:calibration} provides a brief discussion of calibration methods, in particular for data-driven ones.

\section{Observation and Modeling of Space-Charge Effects}\label{sec:history}
This section contains a review of relevant observations of space-charge effects, and of how they led to development in their understanding and description.

\subsection{Calorimeters}

\subsubsection{The NA48 Liquid Krypton Calorimeter and the One-Dimensional Model}\label{sec:NA48}
Detailed discussion of space charge effects in ionization detectors came first from studies and operation of electromagnetic calorimeters.
The NA48 Collaboration 
provided a report of observation of space-charge effects in a quasi-homogeneous liquid-krypton calorimeter, together with an analytical model describing space and time dependence of the effect \cite{Palestini:1998an}. The sensitivity to space charge came from the combination of several conditions: 
\begin{itemize}
\item	The flux to which the detector was exposed was equivalent to a charge density injection of positive ions exceeding 100~pC~cm$^{-3}$s$^{-1}$ in the central part of the calorimeter.
\item	The nearly longitudinal readout structure made the detector response sensitive to the position of the axis of the electromagnetic shower within the readout cell, which had a gap of about 1.0 cm. 
\item	The readout cells were operated at a voltage of only 1.5~kV for the first year of data taking. 
\end{itemize}
In these conditions, space charge effects were identified and corrected for, including their time dependence within the beam spill:  the effect increased during the first 1.6~s of exposure to the beam, corresponding to the time required to reach the equilibrium condition of the density of positive ions, and remained stable during the rest of the spill.
After that, the density of positive ions reduced to zero during the interval to the next spill.
 
 The model developed in \cite{Palestini:1998an} was based on the use of a continuity equation for the density of positive ions:
  \begin{equation}
\frac{~\partial \rho^+}{\partial t} + \frac {\partial(\rho^+ \! v^+_x)}{\partial x}= K \; .  \label{eq:continuity}
\end{equation}
\begin{figure}[t]
\begin{center}
\includegraphics[width=0.7\textwidth]{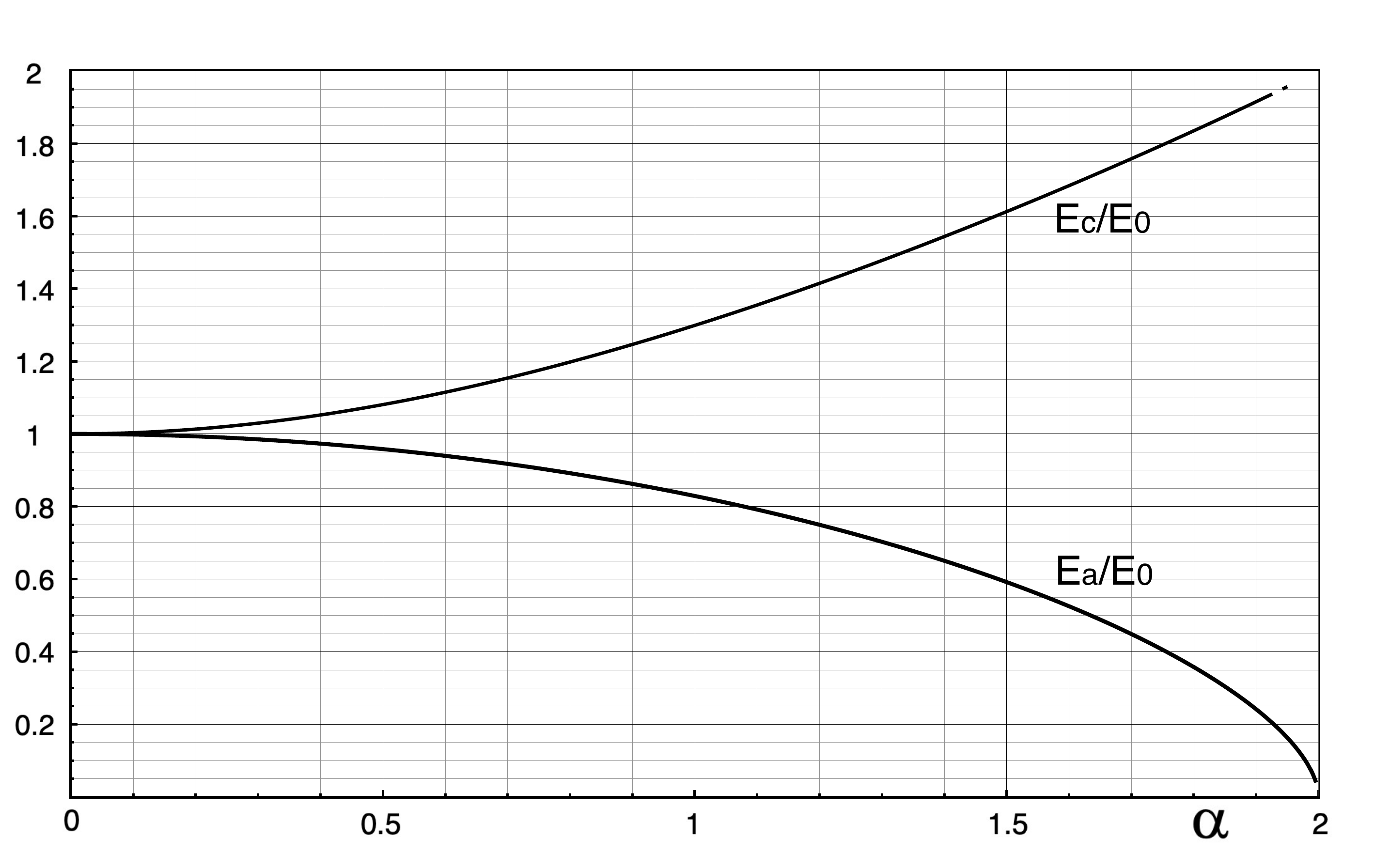}  
\caption {Normalized electric field at the anode $E_\text a / E_\circ$ and cathode $E_\text c / E_\circ$ 
as a function of the dimensionless parameter $\alpha$. Reproduced from reference~\cite{Palestini:1998an}.}
\label{fig:Ea-Ec}
\end{center}
\end{figure}
The detector geometry allows to neglect the dependence on the coordinates parallel to the electrodes.  The parameter $K$ is the injection rate of the charge density of positive ions, which includes the initial recombination of electrons and ions. 
Besides this effect, the presence of the electrons is ignored, since they are removed from the drift cell
in a time about $4 \times 10^5$ times shorter than the ions. 
Under steady conditions, the space charge density satisfies the equation
\begin{equation} 
\rho^+(x)= \frac{K\, x}{\mu^+E(x)}\, ,  \label{eq:rho} 
\end{equation}
where the ion mobility is introduced with $v^+_x = \mu^+ E_x$. 
The electric field is determined by the boundary conditions and by the Gauss equation
\begin{equation}
\frac{dE_x}{dx} =  \frac{~\rho^+}{\epsilon}   \, ,\label{eq:gauss} 
\end{equation}
which is solved as 
\begin{equation}
E_x(x) =  E_\circ \sqrt{(E_\text a/E_\circ)^2 + \alpha^2 (x/L)^2 } \;,  \label{eq:Ex}
\end{equation}
where 
$E_\text a$ is the value 
at the anode, determined by the boundary 
$\int\! E_xdx = V_\circ = E_\circ \,L$,
integrated from anode (at $x=0$) to cathode (at $x=L$), where $V_\circ$ is the voltage difference between anode and cathode, 
and $E_\circ$ is 
the uniform field strength that would be present 
for $K=0$, i.e. for vanishing space charge. 
\begin{figure}[t]
\begin{center}
\includegraphics[width=0.7\textwidth]{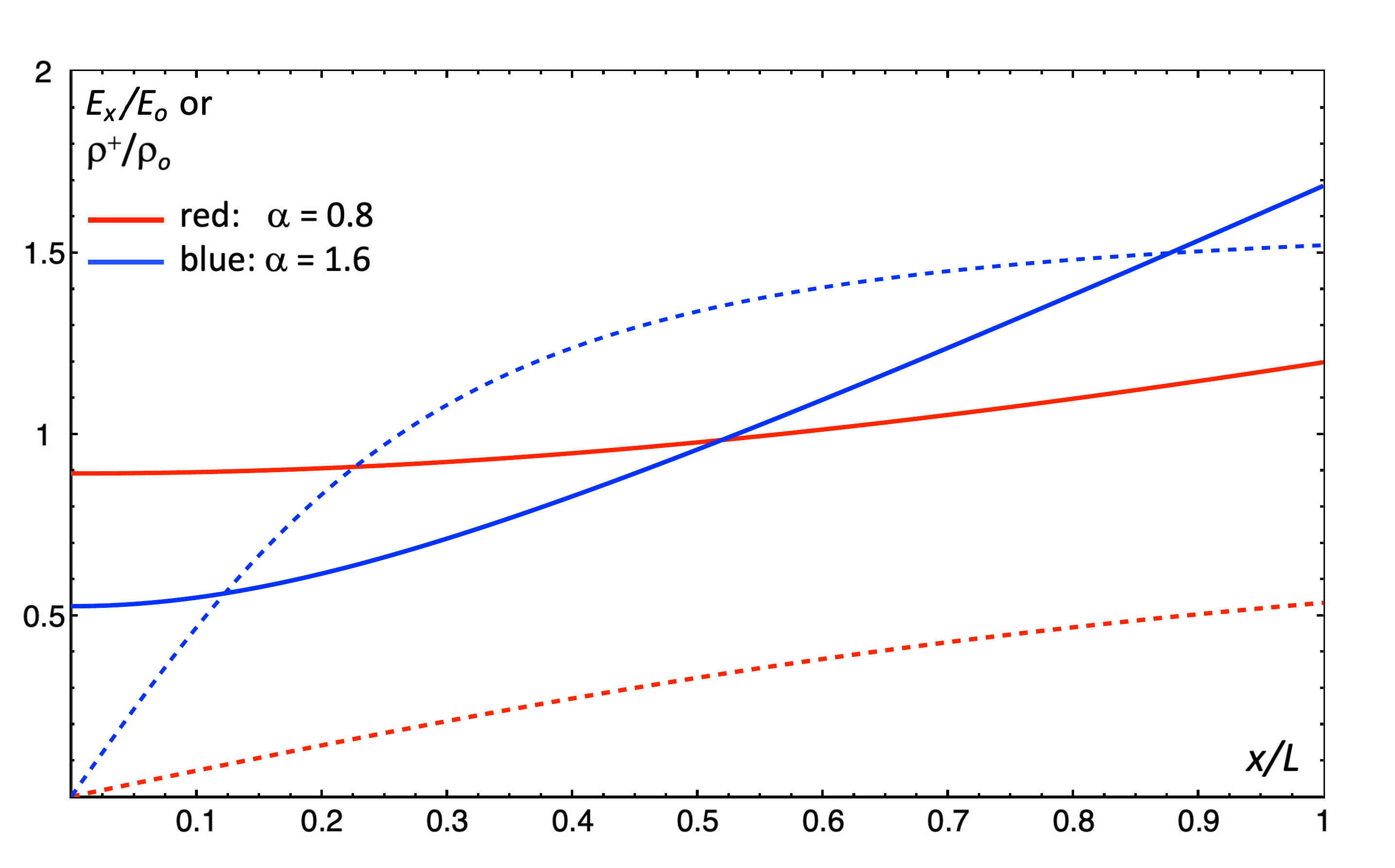}
\caption {Electric field (continuous lines) and charge density (dashed line) behavior for $\alpha = 0.8$ (red) and $\alpha = 1.6$ (blue). The horizontal axis is the drift coordinate divided by the gap length ($x/L$), with $x=0  \,(1)$ at the anode (cathode).  The electric field is in units of $E_\circ=V_\circ/L$, and the charge density in 
units of $\rho_\circ =\epsilon E_\circ/L$. Reproduced from reference~\cite{Palestini&Resnati:2020}.
}
\label{fig:example-E-rho}
\end{center}
\end{figure}
The dimensionless parameter $\alpha$ is defined as 
\begin{equation}
\alpha =  \frac{L}{E_\circ} \sqrt{\frac{K}{\epsilon \mu^+}} \;.  \label{eq:alpha}
\end{equation}
Figure~\ref{fig:Ea-Ec}  shows the dependence of $E_\text a$ and of $E_\text c$, the electric field at the cathode, on the parameter $\alpha$. 
At the anode, the field is lower than 
$E_\circ$, while the opposite holds at the cathode, to a greater extent.   
Figure~\ref{fig:example-E-rho} shows the values of $E_x/E_\circ$ and $\rho^+ / \rho_\circ$ versus $x/L$, for values of the parameter $\alpha$ equal to 0.8 and 1.6.
The density $\rho_\circ$, defined as $\epsilon E_\circ /L$, is the natural scale for the space charge density. 
It is equal to the ratio of the charge density on the surface of the electrodes, when no space charge is present, to the length of the drift gap.  
From Equation~\ref{eq:alpha},  $\alpha^2$  can be described as the charge density injection rate $K$, multiplied by an effective ion drift-time across the gap  $L/(\mu^+ \! E_\circ)$, and divided by  $\rho_\circ$.  
Figure~\ref{fig:Ea-Ec} shows that the effects of space charge are relevant for $\alpha = 0.5$--1, and become large for values exceeding this range.

The observation made by NA48 covered a range of $\alpha$ up to about 1.1, showing good agreement between predictions and observations.

\subsubsection{Developments for Liquid Argon Calorimetry}
Further advancements were driven by interest in the response of liquid argon sampling calorimeters, like those used in the ATLAS detector,
when operated in the high rate environment of the Large Hadron Collider, and in particular for the foreseen upgrade to the High Luminosity Large Hadron Collider (HL-LHC). 
Particular attention was devoted to the study of critical conditions reached at very high intensity.

The occurrence of critical conditions had already been remarked in \cite{Palestini:1998an}, as illustrated in Figure~\ref{fig:Ea-Ec} and Equation~\ref{eq:Ex}:
for $\alpha=2$ the electric field vanishes at the anode, and takes a linear dependence on $x$ as $E_x= 2\, E_\circ (x/L)$. 
For larger values of $\alpha$ it had been suggested that the active region contracts to a gap of length $L' = 2 \, L/\alpha$, detached from the anode by $L-L'$, with 
$E_x=2\, E_\circ (x'/L')$ for $x'=x-(L-L')>0$, while $E_x$ is highly suppressed for  $x \leq L-L'$. 

These conditions of very large space-charge effects were studied with more detailed analytical descriptions \cite{Rutherfoord:2002zr, LArCalorimeters, ion-mobility-critical}, including the charge density of electrons and bulk recombination in the continuity and Gauss equations. 
Laboratory measurements were performed on calorimetric cells exposed to $\beta$ sources. The onset of the critical condition, described by the authors as cell {\em closing down}, was observed and found in substantial agreement with expectations
\cite{ ion-mobility-critical}.  The transition into critical conditions 
was used to define 
the minimum angle to the collision axis at which liquid argon detectors of a given design could be operated at HL-LHC.
Naturally, such limits could be overcome operating the calorimetric cells at higher voltage or, more effectively, with smaller gaps, as the value of the $\alpha$ is proportional to  $L^2/V_\circ$, as shown in Equation~\ref{eq:alpha}.  
Besides, for a given ratio $L^2/V_\circ$, the value of $\alpha$ is affected by the uncertainty in $\mu^+\!$, for which values in the range of  0.8 to 2.0$\times 10^{-7}$~m$^2$s$^{-1}$V$^{-1}$  have been reported (references in ~\cite{Palestini&Resnati:2020}). 

\subsection{Liquid Argon Time Projection Chambers}
Large liquid argon TPCs were proposed as neutrino detectors \cite{LArTPC-Rubbia} as well as for searches of nucleon decays 
\cite{interface-block}.  Such detectors combine the function of tracking devices, with resolution at the scale of 1~mm, determined by diffusion of the drifting electrons, 
together with calorimetric measurement from the collected charge. 
Detectors of this kind have been operated underground, in condition of very low intensity, or near surface.   When operated near surface, the ionization due to cosmic rays corresponds to a charge density injection of approximately $2\times 10^{-10}$~C\,m$^{-3}$s$^{-1}$, and may cause visible effects of space charge. In fact, the six orders of magnitudes between this value and the one quoted in Section~\ref{sec:NA48} are balanced by the increase of the drift length from 1~cm to a few meters, together with a reduction of the average electric field from 1.5 kV~cm$^{-1}$ to typical values of 0.5 kV~cm$^{-1}$.

Compared to the calorimetric cells discussed above, the liquid argon TPCs add additional complexity to the effects of space charge, because: 
\begin{itemize}
\item	These detectors may have lateral dimensions comparable to the drift length, so that border effects are relevant and the description cannot be reduced to one dimension.
\item	Fluid motion may alter the drift of positive ions, moving them away from electrostatic equilibrium, and therefore changing the way they affect  the electric field.
\end{itemize}

Large liquid argon TPCs have been built and operated near surface in ICARUS, MicroBooNE, and, more recently, ProtoDUNE. A brief review of their observations is given in the next three sections.
%

\subsubsection{ICARUS} \label{sec:icarus}
\begin{figure}[h]
\begin{center}
\includegraphics[width=0.7\textwidth]{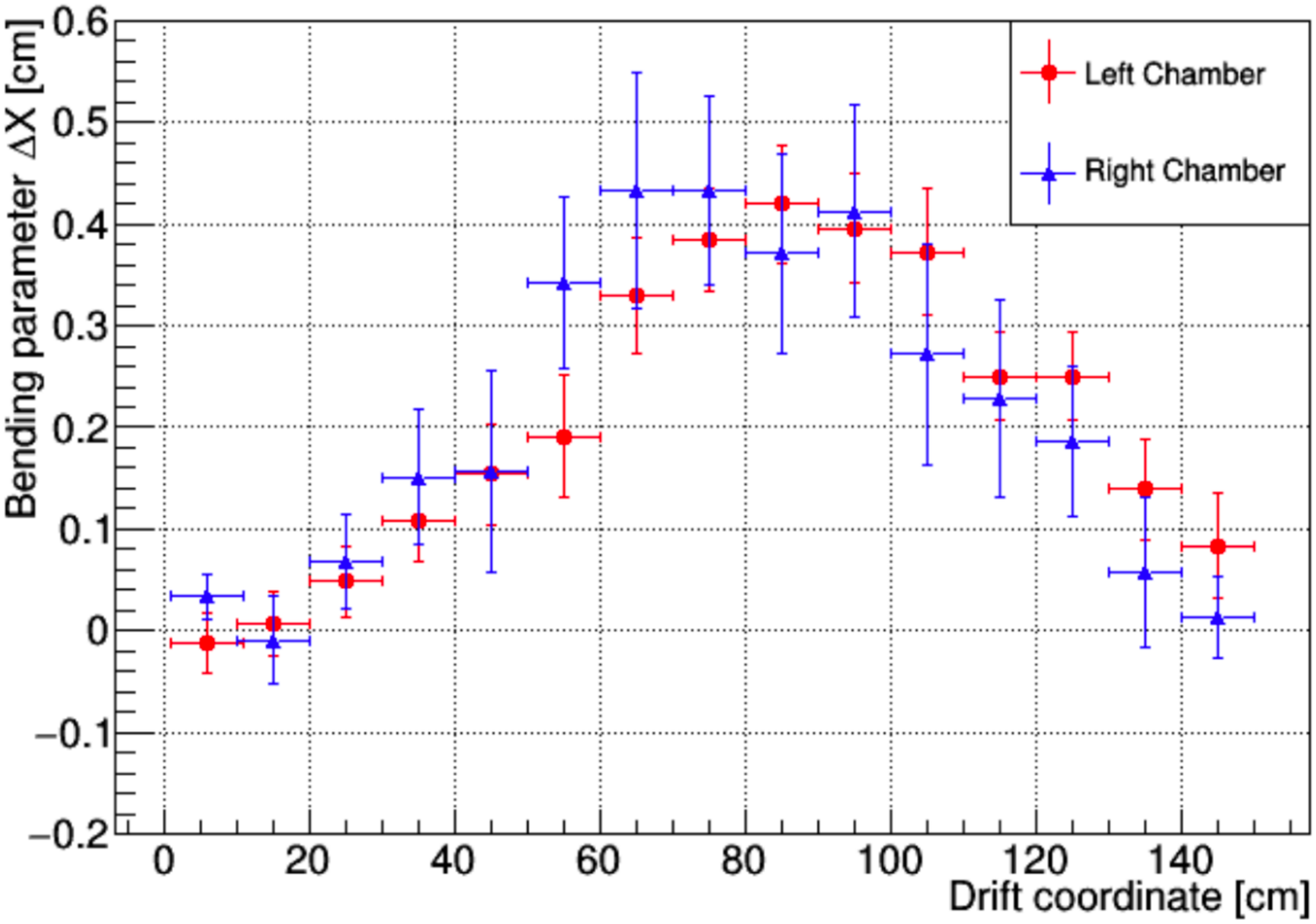}
\caption
{
Bending parameter $\delta x  =v\,^e_\circ \times \delta t (x)$ as a function of the drift coordinate $x$, for the two drift volumes of an ICARUS detector module. Reproduced from reference~\cite{Antonello:2020qht}.
}  
\label{fig:ICARUS-2}
\end{center}
\end{figure}

The ICARUS detector is formed by two modules with a double drift-gap of 1.5~m and transverse active size of  $3.2 \times 18.0$~m$^2$. The detector collected data
 at the Gran Sasso underground laboratory, and on surface, where space-charge effects were studied \cite{Antonello:2020qht}.  
 The observations were analyzed in terms of the delay in the collected charge, due to the distortion to the electric field. 
 The apparent value of the drift coordinate is shifted by 
 \begin{equation}
\delta x(x) =v\,^e_\circ \times \delta t (x)=  \int_0^x \left( \frac{v\,^e_\circ}{v\,^e (x')} - 1\right) dx' = \int_0^x \left( \frac{v\,^e_\circ}{\mu^+E_x(x')} - 1\right) dx'\;,
\end{equation}
where $v\,^e_\circ$ is the electron drift velocity at the average electric field $E_\circ$, and $v\,^e (x')$ is the drift velocity at the field strength $E(x')$ 
present at the drift coordinate $x'$, as in Equation \ref{eq:Ex}.
The authors refer to $\delta t$ and $\delta x$  as {\em bending parameters}. 
\begin{figure}[t]
\begin{center}
\includegraphics[width=0.95\textwidth]{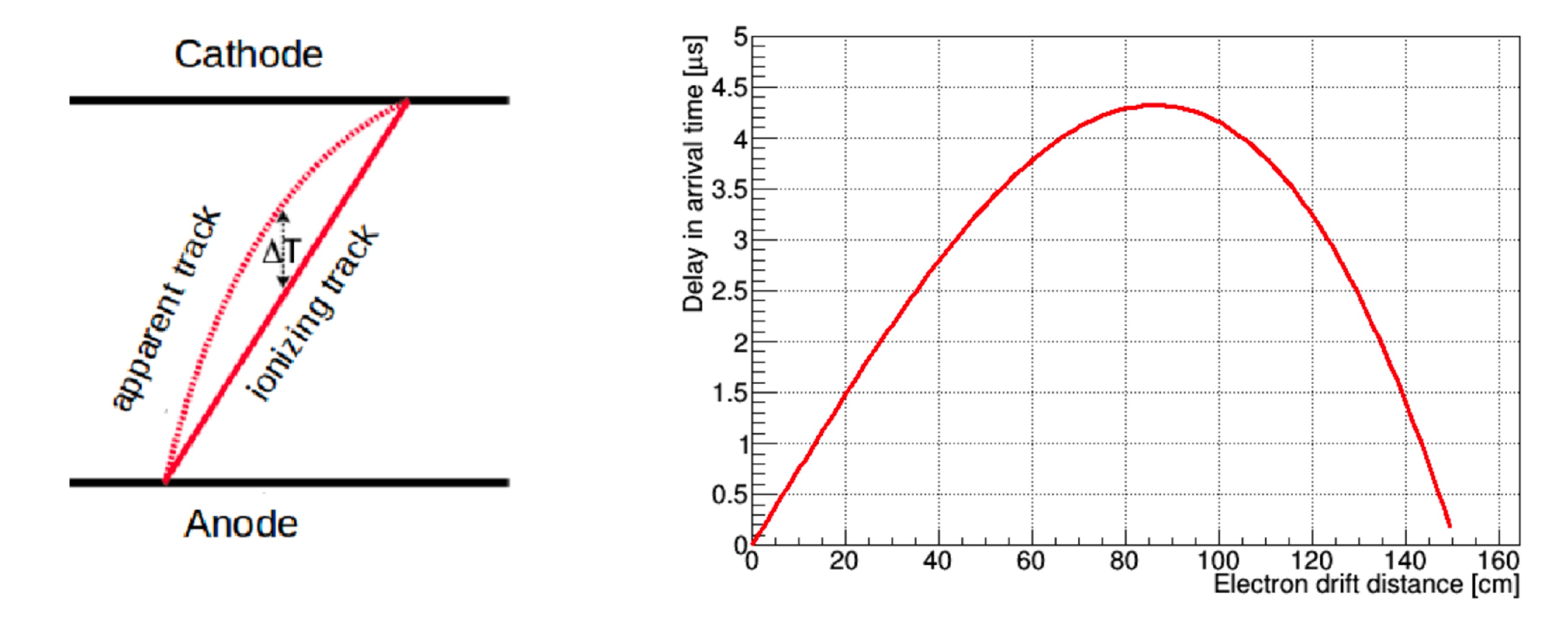}
\caption
{
Left: schematic view of the drift time delay $\delta t$.  Right: drift time delay dependence on the drift coordinate $x$.
The anode is at $x =0$~cm and the cathode at $x=150$~cm. Reproduced from reference~\cite{Antonello:2020qht}. 
}
\label{fig:Fig-ICARUS-1}
\end{center}
\end{figure}
Figure~\ref{fig:Fig-ICARUS-1} illustrates the observable, and shows the expected values of $\delta t (x)$  with $E_\circ=500 \, \text{V cm}^{-1}$, 
for $K=1.9 \times 10^{-10} \text { C\,m}^{-3}\text s^{-1}$, $\mu^+=0.9 \times 10^{-7} \text m^2 \, \text{kV}^{-1}\text s^{-1}$. 
These values correspond to $\alpha = 0.4$. The prediction for the bending parameter includes corrections from modeling of electrodes and field-cage, and the small contribution to space charge from negative ions, due to electron capture from impurities. The result of the measurement is shown in Figure~\ref{fig:ICARUS-2}, which is in good agreement with the expectation, for the quoted values of $K/\mu^+$.

 

\subsubsection{MicroBooNE} \label{sec:MicroBooNE}
\begin{figure}
\begin{center}
\includegraphics[width=0.7\textwidth]{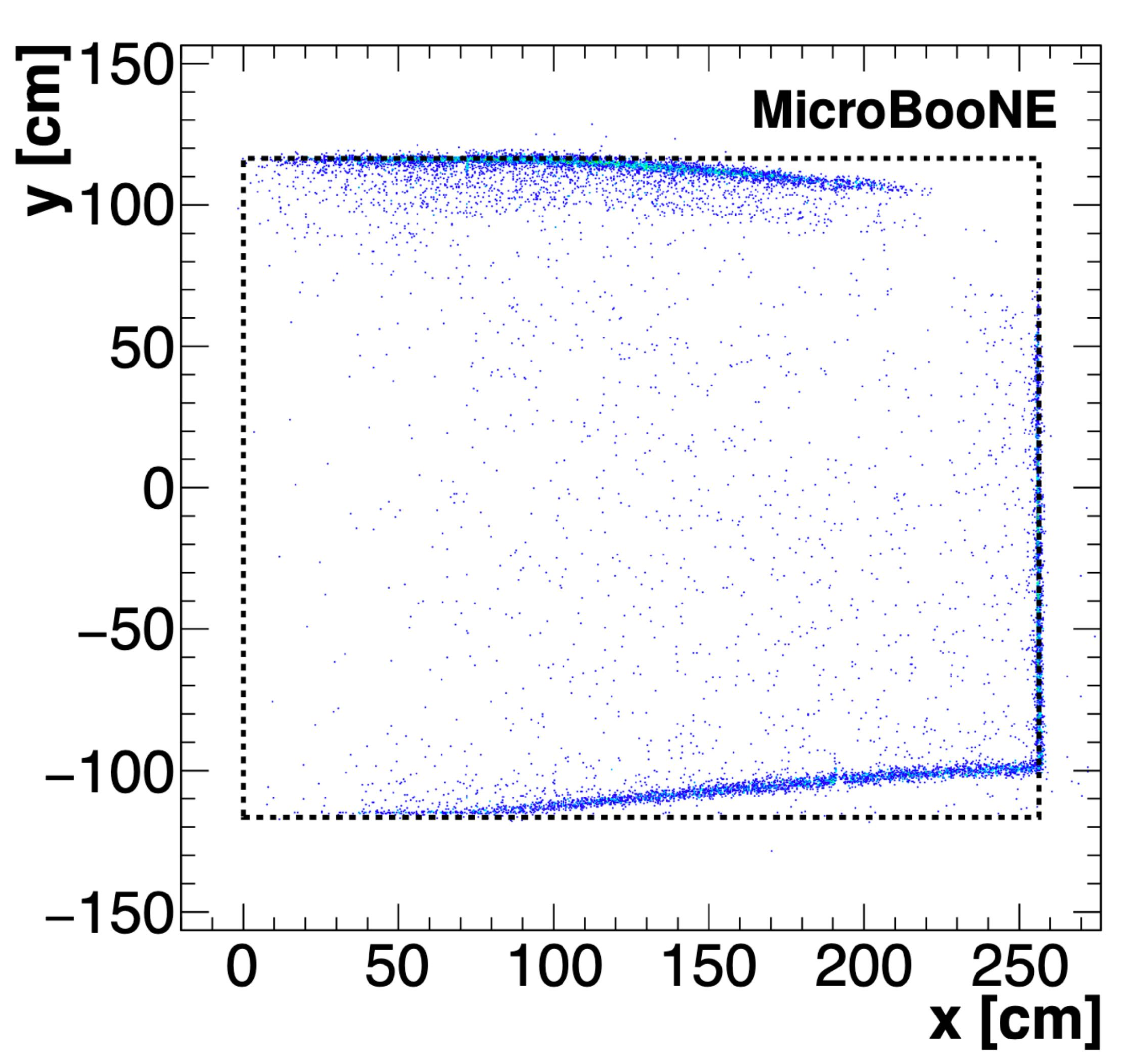}
\caption
{
Entry/exit points of reconstructed cosmic muon tracks in the MicroBooNe detector, projected on the $xy$ plane. The anode is located at $x=0$~cm
and the cathode at $x =256$~cm.  The data points accumulate on the cathode plane and on the profile of the top and bottom side faces, 
which appear distorted towards the center of the detector because of a transverse component in the electric field due to space-charge. Reproduced from reference~\cite{Abratenko:2020bbx}. 
}
\label{fig:MicroBooNE-DeltaY}
\end{center}
\end{figure}

The MicroBooNE liquid argon TPC has collected data from neutrino interactions at the BNB facility of Fermilab. 
The detector has a drift gap $L=2.56$~m (along the $x$ axis, horizontal), transverse size of 2.32~m along the vertical direction ($y$ axis) and
10.37~m along the beam direction ($z$ axis). It has been operated at a nominal electric field of 274~V/cm. 
Under these conditions, and using $\mu^+ = 1.6 \times 10^{-7}$~m$^2$s$^{-1}$V$^{-1}$, 
 the $\alpha$ parameter is approximately equal to 0.81, and space charge effects were clearly observed and extensively reported by the MicroBooNE 
 Collaboration
 ~\cite{Abratenko:2020bbx}:  uncorrected crossing  tracks appear bent in the $xy$ and $xz$ projections more than in ICARUS, with sagittas on the centimeter scale.
 Besides, the most apparent effect of space charge  occurs on the side faces: 
 the transverse coordinates of entry or exit points of ionizing particles crossing the field-cage
are shifted towards the center of the detector. The effect is illustrated in Figure~\ref{fig:MicroBooNE-DeltaY} for tracks crossing the top or bottom side faces of the detector, far from the upstream and downstream side faces, for which  
the apparent $y$ coordinate is plotted against the drift coordinate $x$. The distortion vanishes at the anode ($x=0$) and it is maximum for electrons that have drifted all the distance from the cathode.  The distortion of the $x$ coordinate of the end-point is expected to be very small, since near the field-cage the field is constrained to $E_x =E_\circ$,  apart from minor effects due granularity of the field-cage electrodes. 
The transverse distortions due to space charge are discussed further in Section~\ref{sec:boundary-effects}. 

MicroBooNE uses calibration procedures based on an ultraviolet laser system and on a data-driven approach, which exploits cosmic muons and the limited distortion of track end-points, as discussed below in Section \ref{sec:calibration}.

\subsubsection{ProtoDUNE Single-Phase Detector} \label{sec:ProtoDUNE}
\begin{figure}[t]
\begin{center}
\includegraphics[width=0.95\textwidth]{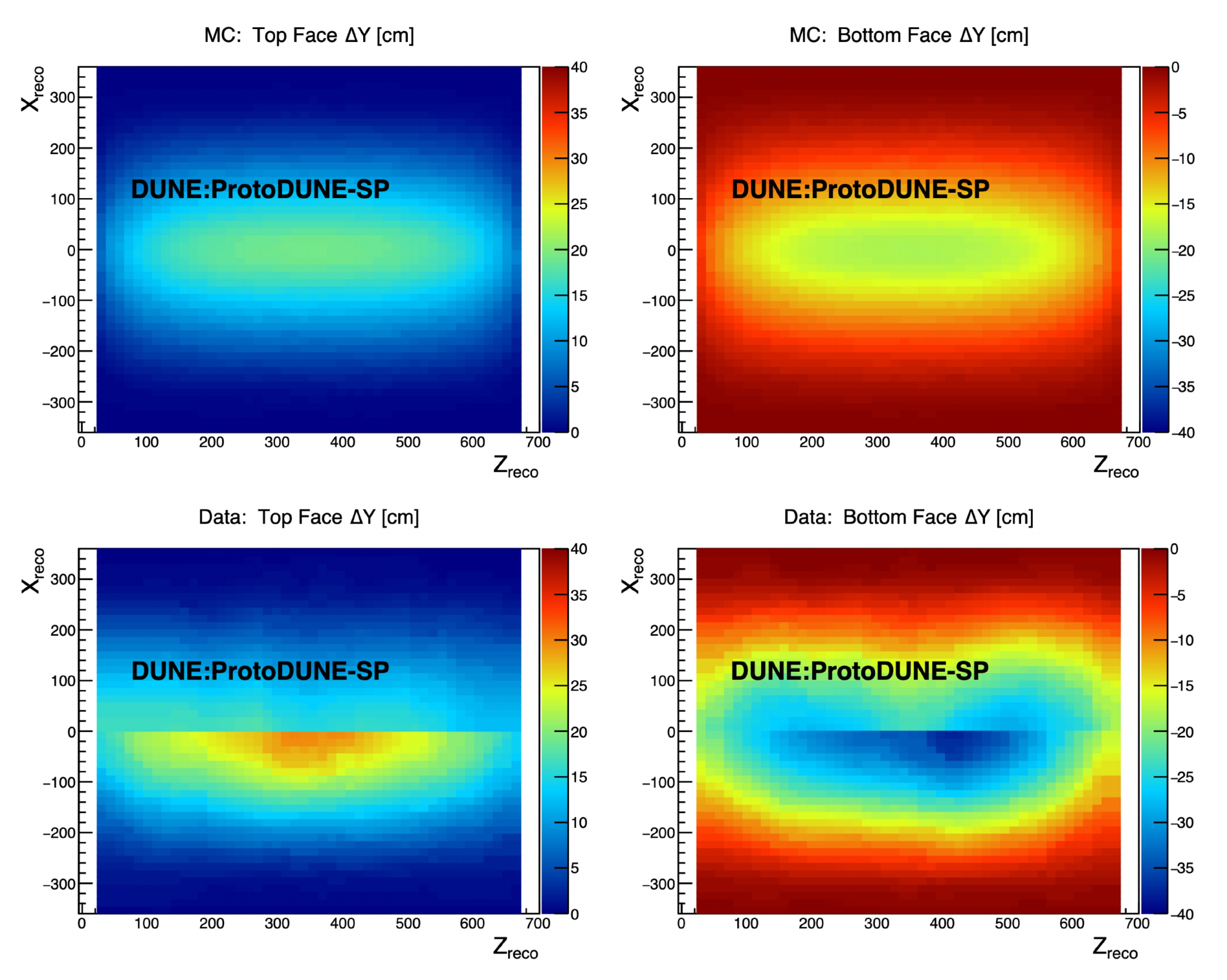}
\caption
{
Predicted and observed transverse distortion on the top and bottom side faces, in the ProtoDUNE Single-Phase detector. 
The top panels show the distortion 
as predicted by an approximate numerical model.   
The amplitude of the distortion is maximal near the cathode ($x=0$), where it reaches about 20~cm, and vanishes on the boundaries with the anodes ($x=\pm \, 360$~cm) and with the other side faces. The observed distortion, bottom panels, is qualitatively comparable, but reaches larger values and shows large asymmetries between the top and bottom faces, and between the two sides of the cathode surface.
Reproduced from reference~\cite{ProtoDUNE-SP-performance}.
}
\label{fig:ProtoDUNE}
\end{center}
\end{figure}
The ProtoDUNE Single-Phase LAr TPC has collected data at the CERN Neutrino Platform.  The detector has an active volume of $7.2 \times 6.0 \times 6.9 \, \text m ^3$,  
with a cathode plane separating two regions with drift gaps of 3.6 m.   Operated at 500~V\,cm$^{-1}$, the parameter $\alpha$ is about 20\% smaller than in MicroBooNE, 
but space charge effects are expected to be larger, because spatial distortion scales approximately as $\alpha \, L$.
Figure ~\ref{fig:ProtoDUNE} \cite{ProtoDUNE-SP-performance}  shows the transverse distortions to the tracks end-points, when they cross the top or bottom face of the active volume. The drift occurs horizontally, along $x$, with the cathode at $x=0$ in these plots. The distortion is directed vertically, along $y$. 
The top panels show the results of a numerical model that provides an approximate description of the effects of space charge and of the boundary condition imposed by the field cage.  
The bottom ones show the corresponding observation. 
Some features of the model, such as larger effects near the cathode and no effects on the boundaries with the anodes and the lateral faces, 
are visible in data, but they come together with large changes in the amplitude of the distortion and with large asymmetries. 
A similar situation occurs near the lateral side faces of the detector, where the end-points are shifted along the $z$ direction.  

Fluid motion, induced by thermal gradients and by recirculation of liquid argon, is likely to be a main contributor to the differences between the model and the observation, since its range in velocity, from fraction of mm to several cm per second, exceeds the drift velocity of positive ions, which is equal to a few mm per second.
The large asymmetries between the two sides of the cathode observed in ProtoDUNE may be related to the construction of the electrode as a closed surface, and to the asymmetric location of the inlet/outlet ports for liquid argon circulation.

\section{Space Charge Beyond the Basic One-Dimensional Model}\label{sec:models}
Space charge affects the electric field and may causes significant deviations from the behavior of an ideal TPC. 
As seen in the examples discussed above, the distortions caused by space charge may exceed by two order of magnitude the 
resolution limit of about 1~mm imposed by electron diffusion. 
Besides, particle identification by $dQ/dX$ is affected by local variations of the ionization yield, which depends on the electric field
through initial recombination, 
and by changes in the absolute length scales over which the charge yield is measured, which varies with the gradients of the spatial distortion.

A discussion of numerical results and analytic approximations has been recently presented in reference~\cite{Palestini&Resnati:2020}. We refer to that analysis for the discussion of local variation in the ionization yield and in the length scale, and for the effect of a negative ions, formed in electron capture by impurities.
 The following sections deal with effects related to the side faces and to the detector aspect ratio, i.e.\ the relative size of the drift gap and of the detector lateral extension. The term {\em longitudinal} is used for the effects that occur along the nominal drift direction, denoted as $x$, while the term {\em transverse} is used for the effects along the $y$ and $z$ directions.  The anode is at $x=0$, and one detector side face, namely a face of the field cage, is at $y=0$.
  


\subsection{Side Faces and Field Cage} \label{sec:fc-effect}
While calorimetric cells might be described with the one-dimensional model of Section~\ref{sec:NA48}, TPC detectors needs a treatment in more dimensions, which cannot be described with analytical equation as simple as equations \ref{eq:rho} and \ref{eq:Ex}.
As the number of dimensions increases, additional care must be taken in dealing with boundary conditions.
For example, when making the approximate assumption on the space-charge density  $\rho^+ \simeq K\, x /(\mu^+ E_\circ)$ and using it as source for a perturbative treatment $E = E_\circ + E^\prime$, the term $E^\prime$ cannot be solved using $\nabla \cdot  E^\prime \simeq \rho^+ \!/\epsilon$ together with empty-space boundary conditions. 
The problem of such approach is that electrodes and field cage react to the presence of $\rho^+\!$, effectively producing a distribution of image charge of comparable density.\footnote{
As an illustration, consider for simplicity the one-dimensional model at lowest level in $\alpha^2$: for a space charge of integrated density  $\Sigma^+=\int \rho^+ dx$, the cathode reacts with an increase in the absolute value of its surface density by $(2/3)\,\Sigma^+$, and the anode with a reduction by $(1/3)\,\Sigma^+$.}

\begin{figure}[ht]
\begin{center}
\includegraphics[width=0.93\textwidth]{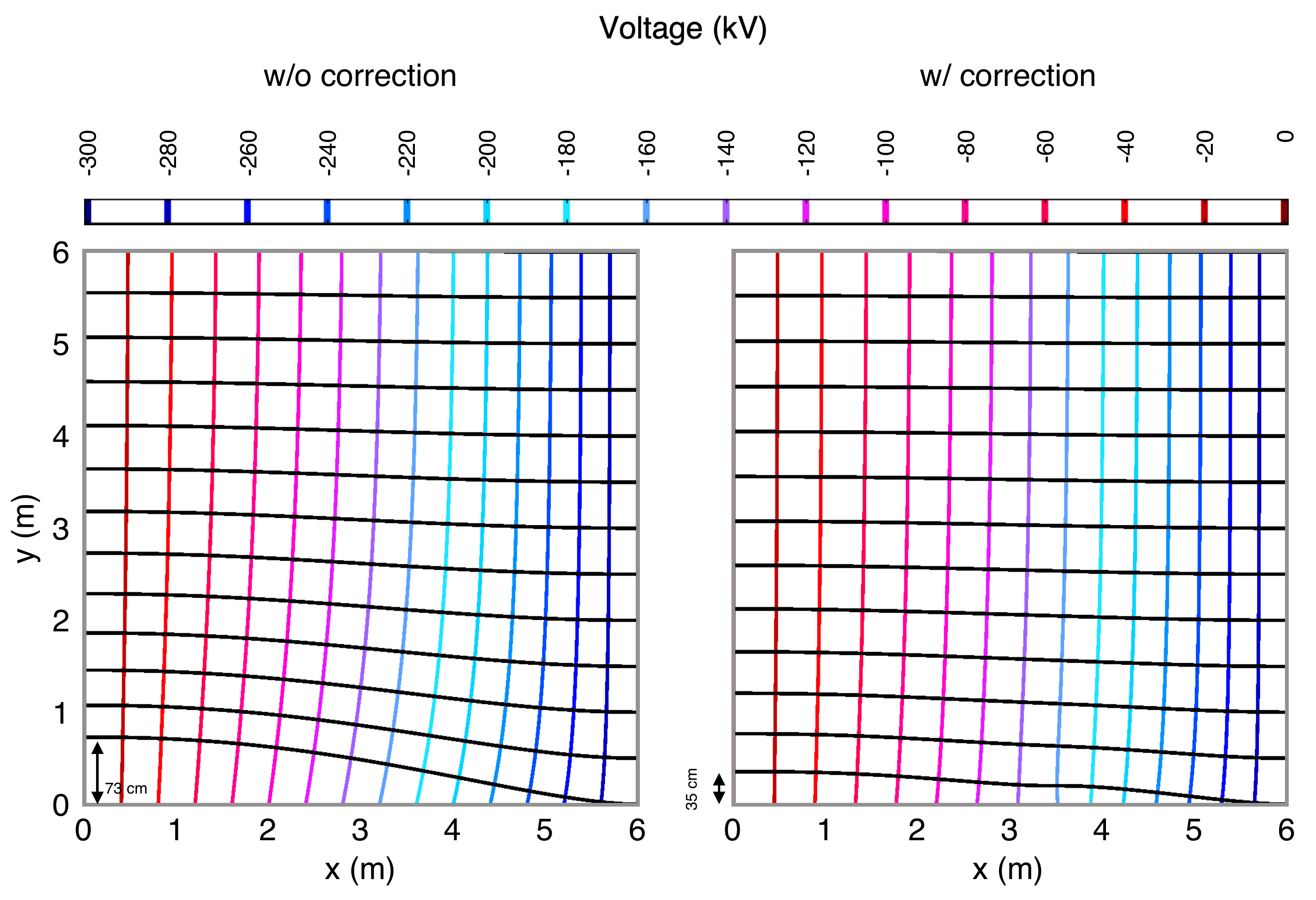}
\caption
{
Contours of equal voltage V(x,y) and drift paths, with a usual voltage gradient at the field cage
(left), and with a third voltage connection at x = 3.5 m, as discussed in the text (right). The anode is at $x=0$, the cathode at $x=6$~m, and the field cage 
at $y=0$. Reproduced from reference \cite{Palestini&Resnati:2020}.
}
\label{fig:2D-V-paths}
\end{center}
\end{figure}

For this reason, describing the inward distortion of sections \ref{sec:MicroBooNE} and \ref{sec:ProtoDUNE} 
just as the result of the attraction from positive space charge in the detector volume, is a simplification which may preclude a wider comprehension of the situation.  
To this purpose, consider Figure~\ref{fig:2D-V-paths}, where a TPC with $L=6$~m and much wider lateral extension is studied with a numerical evaluation of $\rho^+$ and of the components of $E$. 
A cross section in the $xy$ plane is shown, near the side face and far from the boundary in $z$, drawing equipotential contours and drift paths.  
The anode and cathode are at $x=0$ and 6~m, and the field cage is at $y=0$. The space charge density corresponds  to $\alpha=1.15$.   
On the plot on the left, the boundary condition is a linear 
voltage distribution along the field cage: $V_\text {fc}=- V_\circ x/L$.  
Far from the field cage, for $y \gtrsim L$, the equipotential surfaces are parallel, at a distance to each other that depends on $x$ following the pattern of equation~\ref{eq:Ex}.  As the boundary $y=0$ is approached, the mismatch with the voltage gradient imposed by the field cage
causes the equipotential surfaces to bend, generating a transverse component of the field. Hence, the drifting electrons follow curved paths, with the maximum distortion equal to 73~cm for electrons drifting from $x=6$~m, $y=0$~m. 

If the voltage $V_\text {fc}(x)$ is instead modified to follow the behavior of  $V(x) \!=\!-\! \int \! E_x dx$ at $y \ge L$, 
the equipotential surfaces remain parallel all the way to $y=0$, without any transverse component of the electric field.  
Following reference \cite{Palestini&Resnati:2020}, the plot on the right in Figure~\ref{fig:2D-V-paths} shows the result of a very simple modification to the field cage:
 the slope of $V_\text {fc}$ changes at $x=3.5$~m, where $V_\text {fc}$ is given the value $V(x,y\ge L)$ corresponding at the same $x$ but far inside the detector. $V_\text{fc}(x)$ still follows a linear behavior, with different slopes, between this point and the electrodes. 
 As a result, the electric field $E_x(x,y\!=\!0)$ imposed along the field cage 
 is much closer to $E_x(x,y\! \gtrsim \!L)$, the transverse component $E_y$ is suppressed for  $x \simeq 3.5$~m, and the transverse distortions across the full drift length $L$ are reduced by a factor 2. 
 The correction to the field cage could be tuned to larger values, producing an outward distortion of the drift paths, despite the force due to the space-charge density $\rho^+$ remains attractive.
 
Indeed, the irreducible effect of space charge is the distortion on the longitudinal electric field $E_x$, which at $x=L/2$ and far from the side faces corresponds to a voltage variation  equal to $\alpha^2 E_\circ L/16$, at lowest order in $\alpha^2$. 
The drift delay $\delta t$ and the corresponding longitudinal distortion $\delta x$ receive contributions of opposite sign from the regions close to the cathode
(where $E_x >E_\circ$) and to the anode  ($E_x < E_\circ$), and the effects are further reduced by the partial saturation of the electron drift velocity for typical values of $E_\circ$. 
On the other hand, the transverse distortion

\begin{equation} 
\delta y(x,y,z) = \int_{(x,y,z)}^{x=0} \frac{E_y}{E_x}dx 
\end{equation}

receives in general coherent contributions along the entire drift path. With the usual, linear field-cage voltage profile, the maximum magnitude of
$\delta y$ near the side face is significantly larger than the maximum longitudinal distortion in the central region of the detector.
Moving away from the field cage, the amplitude of the transverse component of the electric field and of the transverse distortion decreases approximately as $\exp[-y/(0.5\,L)]$ \cite{Palestini&Resnati:2020}, and the main manifestation of space charge is the longitudinal distortion.

\subsection{Detector Aspect Ratio} 
 \label{sec:boundary-effects}

The Gauss equation can be written as
 
\begin{equation}
\frac{\partial E_x}{\partial x} =  \frac{~\rho^+}{\epsilon}  -  \frac{\partial E_y}{\partial y} -\frac{\partial E_z}{\partial z}  \, , \label{eq:gauss-3D} 
\end{equation}

where the derivative of the main component $E_x$ is related to the density of the space charge and to the derivatives of the transverse components
$E_y$, $E_z$.
As discussed in Section~\ref{sec:fc-effect}, in usual detector configurations the transverse components  and their derivatives are due to space charge,
and since they are directed outward and their magnitude decreases moving inside the detector, they reduce the effect of $\rho^+\!$ on $E_x$.

In detectors with lateral size at least as large as twice the drift length,  in both transverse directions,
the derivatives of the transverse components of the electric field can be neglected in the central region of the detector, 
defined here as the region far from the side faces by a distance greater than $L$, but still extending from the anode to the cathode. 
In this region the one-dimensional description of space charge is still applicable. 
This is close to the situation reported by ICARUS \cite{Antonello:2020qht}, where only small corrections are needed because of the limited lateral extension of the 
detector.

For narrower detectors, $E_y$ and $E_z$ still vanish in the center of the detector because of contributions of opposite sign from opposite faces, 
but their derivatives collect same-sign contributions from all sides. In this condition, the voltage gradient established by the field cage is able to reduce the effects of space charge across the entire detector volume. 
Naturally, narrow detectors are affected by transverse distortion over a large fraction of their volume, in a way similar to pincushion aberration, maximal near the center of the side faces, and vanishing at the corners and at the center of the detector.

In such condition,  the lateral distortion may be reduced with a modification of the voltage profile at the field cage, 
as discussed above in Section~\ref{sec:fc-effect}, 
but the advantage of such option should be weighed against a corresponding increase in the longitudinal distortion.  Numerical examples can be found in reference \cite{Palestini&Resnati:2020}.

\subsection{Comparison of models and observations}
A discussed above in section~\ref{sec:icarus}, the observation of space-charge effects made by the ICARUS Collaboration \cite{Antonello:2020qht}
has been compared by the authors to the prediction of the one-dimensional model.
The transverse distortions due to the limited vertical extension of the drift volume are of low relevance because the tracks crossing the region near the top and bottom field cage are excluded from the test sample, unless they are also close to the anode. The prediction for the amplitude of the observed longitudinal ($\delta x$) distortion is corrected by about $-6\%$ for the detector geometry and $-12\%$ for the presence of negative ions from capture of the drifting electrons. The procedure is consistent with the considerations and the examples provided in reference~\cite{Palestini&Resnati:2020}. 
In that study, for a detector with lateral extension twice as large as the drift gap along one direction, and much larger along the other, 
the field cage changes the amplitude of the longitudinal distortion near the center by about $-9\%$. A reasonable agreement is also found for the effect of negative ions, where the approximate description provided in \cite{Palestini&Resnati:2020} causes a further change by about $-15\%$ when the electron lifetime measured in ICARUS is used.

The amplitude of the longitudinal distortion observed by ICARUS is used to obtain the ratio of charge injection rate $K$ to the ion mobility $\mu$, with the latter assumed equal for positive and negative ions. With $K$ set at the value of $1.9 \times 10^{-10} \text { C\,m}^{-3}\text s^{-1}$, as  usually done for detectors on the Earth surface and operated at an electric field of 500~V\,cm$^{-1}$, the ion mobility is $\mu\simeq 0.9 \times 10^{-7} \text m^2 \, \text{kV}^{-1}\text s^{-1}$, a value on the low side of the range of the available measurements.

The MicroBooNE Collaboration has provided an extensive report on transverse and longitudinal distortions caused by space charge~\cite{Abratenko:2020bbx}.
As mentioned above in section~\ref{sec:MicroBooNE}, the vertical extension of the detector (along $y$) is similar to the drift gap (along $x$), 
while the extension along the 
third coordinate ($z$) is much larger. Far from the detector boundaries in $z$, the effects of space charge can be predicted with the two-dimensional model described in \cite{Palestini&Resnati:2020}. 
Considering first the transverse distortion, its maximum value is given by $|\delta y_\text{max} |= 0.105\,\alpha^2 L $ for a detector with transverse size much larger than the drift gap $L$.
For the aspect ratio of MicroBooNE, namely with $L=2.56$~m and the opposite faces of the field cage separated by 2.32~m, the effect is reduced by 7\%.  From the observed value $|\delta y_\text{max} | \simeq 23$~cm, 
the value of the space charge dimensionless parameter is determined as $\alpha \simeq 1.0$. 
Near the Earth surface, and with the value of the charge injection rate expected for an electric field of 274 V/cm, the corresponding mobility is $\mu^+\!\simeq 1.1 \times 10^{-7} \text m^2 \, \text{kV}^{-1}\text s^{-1}$.

For the longitudinal distortion, obtained after the calibration procedure described in section~\ref{sec:calibration}, the MicroBooNE collaboration has observed 
in the $xy$ plane, far from the detector boundaries on the $z$ axis, a maximum value $\delta x_\text{max} \simeq 3.5$~cm. 
According to reference \cite{Palestini&Resnati:2020} $\delta x_\text{max} = 0.064\,\alpha^2\gamma\, L$, for a detector
with transverse size much larger than the drift gap $L$, with $\gamma = (dv^e\!/dE)(E/v^e)$ describing the dependence of the electron drift velocity on the electric field.  The distortion is reduced by about 50\% for a detector with the aspect ratio of MicroBooNE.  
The value of the space charge parameter corresponding to the observed $\delta x_\text{max}$ is $\alpha \simeq 0.8$, 
and the corresponding ion mobility is $\mu^+\! \simeq 1.5 \times 10^{-7} \text m^2 \, \text{kV}^{-1}\text s^{-1}$.

It is possible that the different values of the ion mobilities reported here reflect an incomplete description of the physical effects. 
The value obtained by ICARUS is in reasonable agreement with the lower value derived from MicroBooNE, but they refer to different observables. MicroBooNE reports asymmetries in the pattern of the transverse distortion between the top and bottom side faces, and between the opposite faces at the far sides of the detector. The vertical distortion on the top face shows an unexpected dependence on the $z$ coordinate. These observations suggest that the distribution of space charge is affected by fluid convective motion in a significant way --- although not at the level apparently found with the ProtoDUNE Single-Phase detector. In order to explain the difference in the mobility values derived from MicroBooNE, fluid motion should effectively enhance the effect of space charge near the field cage and/or reduce it in the central region of the detector.

\section{Dual-Phase Devices}\label{sec:dual-phase}
In dual-phase TPCs, at the end of the drift regions the electrons cross an extraction grid before entering the vapor phase, where they undergo gas multiplication. 
Part of the positive ions produced in the multiplication process are driven back to the vapor-liquid interface, and a fraction of them may cross the extraction grid and contribute to the space charge density together with the ions from primary ionization. Large dual-phase liquid argon TPCs have been constructed \cite{dual-phase-WA105},
and the subject of positive ions transport at  the vapor-liquid interface has been considered, with different conclusions, in \cite{interface-block} and \cite{interface-favor}.  
While so far positive ions from gas multiplication have not been directly  observed, 
their contribution to space-charge effects in the drift volume has been discussed in \cite{Palestini&Resnati:2020}.
In this section we illustrate the extension of the one-dimensional analysis that includes positive ions from primary ionization and {\em feedback} ions from multiplication in gas.  

The extraction grid is designed to be transparent to drifting electrons and to facilitate their transport into the vapor phase.  
Ignoring electron capture, the flux of electrons at the
extraction grid is $J^e=K\,L$, i.e.\ it is equal to the rate of charge density injection integrated over the drift length. The corresponding flux of positive ions 
entering the drift region is  $J^+\!=\beta \, K\, L$, where the parameter $\beta$ is the product of different factors: 
(a) the fraction of electrons driven to the amplification region; 
(b) the gas multiplication gain; 
(c) the efficiency in driving the positive ions to the vapor-liquid interface; 
(d)  the transport into the liquid; 
and (e) the limited transparency of the extraction grid for positive ions.  
The value of $J^+$ at $x\!=\!0$ is the boundary condition of the continuity equation 
(\ref{eq:continuity}). The stationary solution for $\rho^+$ is 

\begin{equation}
\rho^+\!(x) = \frac{K(x+\beta \, L)}{\mu^+ E_x(x)} \, .
\end{equation}

The Gauss equation (\ref{eq:gauss}) can be directly integrated, obtaining: 

\begin{equation}
E_x(x) =  E_\circ \sqrt{(E_\text a/E_\circ)^2 + \alpha^2 [(x/L)^2 + 2\, \beta \,(x/L)]} \;.
\end{equation}

\begin{figure}[tb]
\begin{center}
\includegraphics[width=0.7\textwidth]{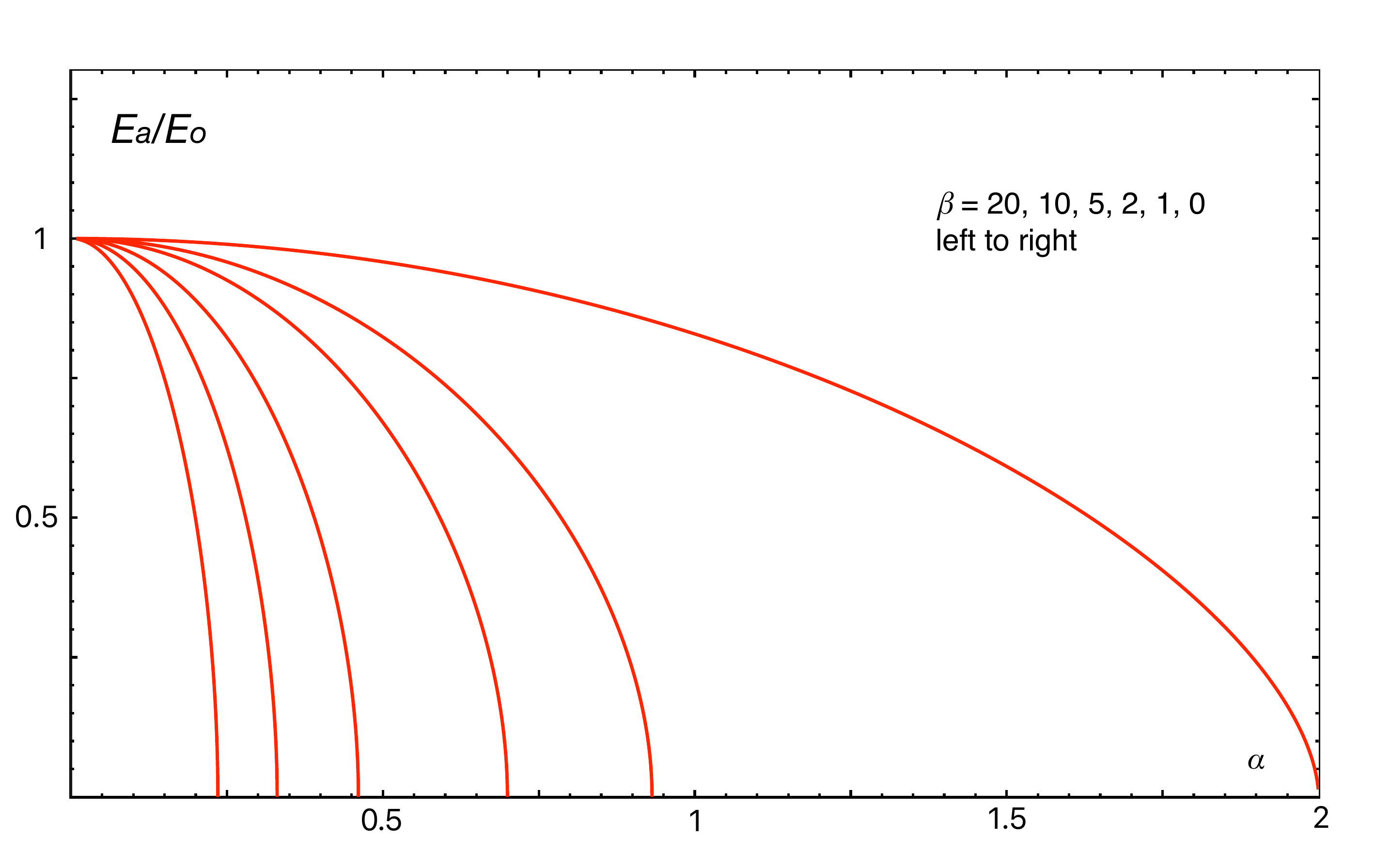}
\caption {Behavior of $E_\text a/E_\circ$ vs.\ $\alpha$, for different values of the feedback parameter $\beta$.
Reproduced from reference \cite{Palestini&Resnati:2020}.
}
\label{fig:backflow1Dcritical}
\end{center}
\end{figure}

In the comparison with equation~\ref{eq:Ex}, we see that the solution is now determined by two dimensionless parameters: the space charge parameter 
$\alpha$ defined in equation~\ref{eq:alpha} and the ion feedback parameter $\beta$.  The values of the electric field at the anode $E_a$ is an integration constant determined as usual by $\int\! E_xdx = V_\circ$.  
The reduction in $E_\text a$ as a function of $\alpha$ is significantly enhanced by the presence of ion feedback, 
as shown in Figure~\ref{fig:backflow1Dcritical}. 
For $\beta \! \ge \! 1$, the critical condition of vanishing $E_\text a$  is reached for values $\alpha \! < \! 1$.

\section{Calibration Methods} \label{sec:calibration}
Calibration procedures are necessary because of the large uncertainties in the value of $\mu^+$, and for uncertainties and possible space dependence in the value of $K$.  Both parameters affect the value of $\alpha$. Besides, calibration may be necessary because of uncertainties in the way fluid motion affects the distribution of space charge.
 
Laser beams for calibration purposes have been used in MicroBooNe \cite{MB-laser-calibration} and design is underway for DUNE
\cite{DUNE-TDR-4}.  Two methods are usually considered. In the first, the laser is used to provide well defined sources of photoelectrons at predetermined locations. For example, photo targets placed on the cathode respond to laser pulses providing a precise measurement of the total drift time and of the transverse distortion for drift paths across the full gap. In the second method, UV lasers are used to generated ionization {\it tracks} in liquid argon TPCs via multi-photon ionization \cite{laser-ionization}. Movable mirrors can be used to steer laser beams across different regions of the drift volume,
suitable for comparison with uncalibrated tracks from reconstruction. 
In MicroBooNE, an iterative correction procedure is used to determine the spatial displacement between true and apparent coordinates, using laser systems placed on opposite side faces of the detector. Alternatively, a calibration procedure can be applied to the crossing points of tracks originating from the two laser beams: 
the expected (true) coordinates of the crossing point can be compared to the corresponding values from uncalibrated tracks, providing directly a local 
measurement of all components of the spatial distortion.

Besides the laser system, MicroBooNE has extended the method of crossing tracks, replacing laser beams with crossing cosmic muons \cite{Abratenko:2020bbx}. 
The method relies on the position of the end-points of tracks crossing the detector boundaries. As discussed in Section~\ref{sec:MicroBooNE},
on the side faces of the detector the spatial distortion is normal to the surface and it can be directly measured on observed tracks, before calibration, while no distortion is present for the end-points at the anode.  
Once all {\em true} end-points are determined, pairs of nearly crossing muons 
can be used to compare the expected coordinate of the (near) intercept point with the corresponding one from uncalibrated tracks.
The calibration of the transverse distortions for electrons drifting from the cathode is not directly available, and MicroBooNE uses an approximation of it,
interpolating the ratio of observed to expected distortion measured on the edges of cathode and side faces. 
In principle, a detector with a double-drift, anode--cathode--anode configuration could avoid such approximation using anode--to--anode crossing muons. 

Deep underground detectors would not benefit from calibration based on cosmic muons, but they should also be free from space-charge effects, and calibration methods are foreseen to reduce instrumental uncertainties of different origin.

\section{Conclusions}
In recent years, the subject of space charge in ionization detectors has seen an increase of interest, driven by the foreseen operation of calorimeters under unprecedented level of radiation, and by the development of large size liquid argon TPCs for neutrino experiment. 
In this paper, the experience gained with two calorimeter technologies and with three TPCs has been reviewed.  The formalism used to describe the main features and the scaling laws of the space charge phenomenology has been discussed along its main lines. A new way to consider the transverse effect near the side faces has been presented. The multidimensional model has been compared to the observations made with the ICARUS and the MicroBooNE detectors.
A brief description of calibration methods has been provided, discussing in particular a data-driven method, that may have further application in the SBN neutrino physics programme which is about to start.  Even with detailed modeling, calibration may remain fundamental for dealing with the effects of fluid motion, which can alter significantly the equilibrium configuration of space charge and electric field. It is possible that modeling of fluid motion will be proven sufficiently accurate, and maybe a space-charge aware detector design, in a wider sense than done so far, will improve the reliability of the predictions. The SBN programme, and further developments on DUNE detector prototypes operated on surface will provide the ground and the motivation for advancement during the next few years.   

\section*{Acknowledgements}
The author thanks Filippo Resnati,  Flavio Cavanna, Kirk McDonald, Michael Mooney, Francesco Pietropaolo, Stephen Pordes, Tingjun Yang and Bo Yu 
for useful discussions on the subject of this paper.

\end{document}